\documentclass[aps,pre,amsmath,amssymb,preprint]{revtex4-1}
\usepackage{hyperref}

\usepackage{graphicx}
\usepackage{dcolumn}
\usepackage{multirow}

\usepackage[utf8]{inputenc}
\usepackage[english]{babel}

\usepackage{graphicx}
\graphicspath{{./figs/}}
\newlength{\figsize} 

\newcommand{\dt}[1]{\frac{\partial #1}{\partial t}}
\newcommand{\dx}[1]{\frac{\partial #1}{\partial x}}
\newcommand{\dxx}[1]{\frac{\partial^2 #1}{\partial x^2}}

\newcommand{\dxtil}[1]{\frac{\partial #1}{\partial \tilde x}}

\newcommand{\dxxtil}[1]{\frac{\partial^2 #1}{\partial \tilde x^2}}

\newcommand{\re}[1]{#1'}

\renewcommand{\arg}[1]{\mbox{arg} \{#1\}}

\begin{document}

\title{Influence of dielectric layers on estimates of diffusion coefficients and concentrations of ions from impedance spectroscopy}

\date{\today}

\author{Maxim V. \surname{Khazimullin}} 
\email{maxim@anrb.ru}

\author{Yuriy A. \surname{Lebedev}}

\affiliation{Institute of Molecule and Crystal Physics - Subdivision of the Ufa Federal Research Centre of the Russian Academy of Sciences, Prospekt Oktyabrya 151, Ufa, Russia, 450075}

\begin{abstract}
We present the analysis of the impedance spectra for a binary electrolyte confined between blocking electrodes with dielectric layers. 
An expression for the impedance is derived from Poisson-Nernst-Planck equations in the linear approximation taking into account the voltage drop on the dielectric layer. 
The analysis shows, that characteristic features of the frequency dependence of the impedance are determined by the ratio of the Debay length and the effective thickness of the dielectric layer. 
The impact of the dielectric layer is especially strong in the case of high concentrated electrolytes, where the Debay length is small and thus comparable to the effective thickness of the dielectric layer. 
To verify the model, measurements of the impedance spectra and transient currents in a liquid crystal 4-n-pentyl-4'-cyanobiphenyl (5CB) confined between polymer-coated electrodes in cells of different thicknesses are performed. 
The estimates for the diffusion coefficient and ion concentration in 5CB obtained from the analysis of the impedance spectra and the transient currents are consistent and agree with previously reported data. 
We demonstrate that calculations of the ion parameters from the impedance spectra without taking into account the dielectric layer contribution lead in most cases to incorrect results.
Application of the model to analyze violations of the low-frequency impedance scaling and contradictions in the estimates of the ion parameters recently found in some ionic electrolytes are discussed. 

\end{abstract}

\pacs{66.10.-x, 66.10.Ed, 77.84.Nh, 77.80.bj}

\keywords{ionic conductivity, impedance spectroscopy, electrode polarization, diffusion coefficient, liquid crystals}

\maketitle

\section{Introduction}
\label{sec:intro}
Ionic conductors are materials in which the electric charge is mainly transported by ions.
An ionic conductivity is observed in a wide class of materials: ionic glasses and ionic liquids, polymers and polymer electrolytes, hydrogels, electrolyte solutions \cite{Dyre:2000:RMP, Kornyshev:2007:JPCB,Bazant:2004:PRE, Macdonald:2011:JPCC}. 
These materials are of considerable technological interest  \cite {Nazri:2008, Ramos:2011} due to the peculiarities of electrical properties, and are subject to active research.
Nevertheless, a deep understanding of charge transport in ionic conductors is far from being complete \cite{Dyre:2000:RMP, Beunis:2013:COCIS, Sangoro:2011:SM, Wang:2012:PRL}.

In a continuum description of the ionic conductivity, charge carriers are characterized by a set of parameters -- valency, equilibrium concentration and diffusion coefficient  \cite{Bazant:2004:PRE, Kornyshev:2007:JPCB}.
To determine ion parameters an impedance or broad-band dielectric spectroscopy in the frequency range ($10^{-6} - 10^7$)~Hz  is widely used \cite{Kremer:2003,Barsoukov:2005,*Lvovich:2012}.
The method is based on measuring of the electrical current flowing in a sample under a small ac voltage, and results representing in the form of frequency dependent complex quantities, which characterize the electrical response of a medium  \cite{Kremer:2003}, such as an impedance, a complex dielectric constant, a complex conductivity  etc.
These quantities are related to each other by simple relationships, and in the following discussion we will concentrate on the impedance $Z = Z' - i Z''$.

The impedance spectra of various ionic conductors demonstrate universal behavior and obey scaling low in a certain frequency range under variation of temperature, charge carrier concentration, sample geometry, etc. \cite{Dyre:2000:RMP, Sangoro:2008:PRE, Serghei:2009:PRB, Sangoro:2011:SM}.
Decreasing the frequency $f$ of the ac voltage applied to the sample, real part $Z'$ and imaginary part $Z''$ of the impedance increase as $f^{-2}$ and $f^{-1}$, respectively.
In the low-frequency range  $Z'$ goes to a plateau, $Z''$ has two local extrema, and the impedance argument $\arg{Z}$ has a well-defined maximum.
At further decrease in frequency the impedance behavior lost the scaling feature and depends on the additional factors, such as electrode material, ion adsorption at electrodes, ion association/dissociation, etc. \cite{Serghei:2009:PRB, Alexe-Ionescu:2009:JAP, Beunis:2013:COCIS}. 

To describe the frequency dependence of the impedance of ionic conductors various theoretical approaches have been proposed \cite{Dyre:1988:JAP, Feldman:2001:MST, Funke:2006:SSI, Sanabria:2006:PRE, Frischknecht:2014:JCP}, one being based on solutions of the Poisson-Nernst-Planck (PNP) equations (see, e.g., \cite{Jaffe:1952:PR, Macdonald:1953:PR, Bazant:2004:PRE}).
In this approach, it is assumed that under an external electric field ions movement is due to migration and diffusion and is described by the continuity equation.
The spatial distribution of the charges and the local electric field are determined in a self-consistent manner from the Poisson equation.
Although the PNP equations are nonlinear, the frequency dependence of the impedance  can be obtained from the solutions of linearized equations in the approximation of small applied voltages (the Debye-Huckel approximation) \cite{Macdonald:1953:PR, Buck:1969:JEAC, Sorensen:1995:JCSFT, Bazant:2004:PRE}.

In the PNP approach the model for a binary electrolyte with blocking electrodes is often used.
It is assumed, that the electrolyte is globally neutral, contains only one type of positive and negative charges and there is no electric current across the boundaries due to the ion movement or electrochemical reactions.
Macdonald \cite{Macdonald:1953:PR} has solved the linearized PNP equations and derived the expression for the impedance in general case taking into account different mobilities and generation/recombination  of the charge carries.
Further, the expressions for the impedance and the dielectric constants were obtained for simplified versions of the model  \cite{Buck:1969:JEAC, Coelho:1991:JNCS, Sorensen:1995:JCSFT, Barbero:2005:LC}, where fully dissociated charge carries  with the same mobility and valence (symmetrical electrolyte) were assumed.

As it shown in Refs.~\cite{Macdonald:1953:PR, Buck:1969:JEAC, Coelho:1991:JNCS, Sorensen:1995:JCSFT, Barbero:2005:LC}, the frequency dependence of the impedance is determined by polarization resulting from the charge separation in the bulk and at the boundaries of the ionic conductor.
The high-frequency behavior of the impedance is dictated by the polarization of the bulk charges due to the ions oscillations relatively to equilibrium positions.
At low frequencies so called electrode polarization phenomena may occur.
A competition between migration and diffusion of ions induces accumulation of the charges near the electrodes (a diffuse charge) with exponential decay of the concentration into the bulk.
Simplifying the picture, it can be considered as a formation of adjacent to the electrodes diffuse layers with the thickness of the order of the Debye screening length $\lambda_D$.
The oppositely charged diffuse layers provide the macroscopic polarization of the sample \cite{Coelho:1991:JNCS},  dominating in the low-frequency range of the impedance spectra.
Note, that adsorption and electrochemical processes may affect the properties of diffuse layers, that is reflected in experimental impedance spectra  \cite{Macdonald:2011:JPCA, Barbero:2012:JCP}. 
These phenomena can be described on the basis of models with partially blocked electrodes and are out of scope of this work.

Experimental impedance spectra of different ionic conductors in a certain frequency range are well described by the model of a binary electrolyte with blocking electrodes.
It allows to estimate the diffusion coefficient and concentration of the ions by a direct modelling or analysis of extrema of the experimental spectra.
In the latter case, the method is called the electrode polarization analysis, because the extrema are observed at the frequencies where the diffuse layer dynamics determines the frequency dependence of the impedance.
Such approach was based on works \cite{Macdonald:1953:PR, Trukhan:1963:SPSS, Buck:1969:JEAC,  Coelho:1983:RPA, Sorensen:1995:JCSFT} and developed in Refs.~\cite{Klein:2006:JCP, Munar:2011:JNCS}.
In particular, it was shown \cite{Klein:2006:JCP}, that the diffusion coefficient can be directly calculated using the value and position of the dielectric loss tangent maximum.
The method was successfully applied to different kind of ionic conductors -- polymer films \cite{Sorensen:1995:JCSFT}, polymer electrolyte \cite{Klein:2006:JCP, Munar:2011:JNCS}, ionic liquids \cite{Sangoro:2008:PRE}.
However, in some cases, the estimates of the ion parameters essentially deviate from those obtained by stoichiometric calculations of the total ion concentration or determined by other experimental methods.
In particular, in some Li-containing polymer electrolytes, ionic liquids, and nonaqueous salt solutions, the ion concentration obtained from the dielectric spectra analysis and using a pulsed-field gradient nuclear magnetic resonance method can differ by up to $4$ orders of magnitude \cite{Wang:2013:PRE}.
Moreover, the results lead to physically contradictory conclusions, such as reduction of free charge carriers with an increasing in the concentration of salts in the electrolyte.
Liquid crystals (LC) are another example of ionic electrolytes, where inconsistent estimates of the ion parameters based on the impedance or dielectric measurements have been reported.
LC are organic liquids with molecules oriented along a given direction characterized by a unit vector (so called director).
The conductivity of liquid crystals originates mostly from impurity ions with usually unknown composition and concentration.
For different liquid crystals with approximately the same viscosity, the diffusion coefficient of the ions differs by three orders of magnitude, that is interpreted as an existence of free ions with the Stokes radius $R_g \sim 0.1$ nm \cite{Naemura:2003:MCLC},  solvated ions with $R_g \sim 1$~nm  \cite{Bremer:1998:JJAP} and colloidal ions with $R_g \sim (5 - 7)$~nm \cite{Huang:2012:JAP}.
For example, for the liquid crystal 4-n-pentyl-4'-cyanobiphenyl (5CB) the diffusion coefficient of the ions differs within one order of magnitude, as reported in \cite{Murakami:1997:JJAP, Sawada:1999:JJAP:1418, Alexe-Ionescu:2009:JAP, Paula:2012:PRE}. 
Note, that despite the low conductivity values, typically $\sim 10$~nS/m, anisotropy of physical properties of liquid crystals results in numerous peculiar electro-optical and electro-kinetic effects absent in isotropic electrolytes \cite{Peng:2015:PRE, Tovkach:2016:PRE}.
To understand these phenomena quantitative characteristics of the ionic conductivity of liquid crystals are necessary.

Omission of the contribution of the dielectric layers at the electrodes in the analysis of the frequency dependence of the impedance can be one possible reason for the difference in the estimates of ion parameters. 
The impedance measurements of isotropic electrolytes are usually carried out in cells with metal electrodes and the Stern layer or so-called compact layer of adsorbed ions can be formed on the electrode surfaces \cite{Stern:1924, Macdonald:1954:JCP, Murakami:1996:JAP, Martin:2005:JAP, Martin:2009:APL}.
On the other hand, typical experimental studies of liquid crystals involve cells with the electrodes, covered by a thin, $\sim (5-50)$~nm, insulating polymer film. 
To take into account the effect of the polymer films or compact layers in the framework of the binary electrolyte model with blocking electrodes, a nonconducting dielectric layers between the electrodes and electrolyte are often introduced \cite{Trukhan:1963:SPSS, Itskovich:1977:PSSA, Sorensen:1995:JCSFT, Bazant:2004:PRE, Olesen:2010:PRE}.
In 1963 Trukhan \cite{Trukhan:1963:SPSS} has derived the expression for the complex dielectric constant on the basis of the solutions of the linearized PNP equations and has shown, that the position of dielectric loss maximum depends on the properties of the dielectric layers.
However, in the modern analysis of the electrode polarization the dielectric layers were neither considered \cite{Macdonald:1953:PR,  Buck:1969:JEAC, Coelho:1991:JNCS} nor neglected  \cite{Sorensen:1995:JCSFT} in the derivation of the frequency dependence of the impedance. 
In this paper the impact of the dielectric layers on the impedance spectra and the estimations of the ion parameters is analyzed.
The paper is organized as follows.
In Sec.~\ref{sec:theory} the binary symmetric electrolyte model with blocking electrodes is used to derive an expression for the impedance from the solution of the linearized PNP equations with mixed boundary conditions for the electrical potential taking into account a voltage drop across the dielectric layer.
Frequency analysis of the impedance expression in the limit of thin diffuse layer is performed and the relations between the impedance extrema, the characteristic times and the ion parameters are found.
In Sec.~\ref{sec:exp} the measurements and the analysis of the impedance spectra of the liquid crystal 5CB in cells with the electrodes coated by the polymer film are reported.
Using the model considered in Sec.~\ref{sec:theory}, the diffusion coefficient and the concentration of the ions are obtained for the fresh made cells of three different thicknesses and during the cell aging over $3000$~h.
For consistency the ion parameters are determined based on transient current analysis in the low and high voltage limits.
Finally, in Sec.~\ref{sec:disc} the obtained results are discussed and compared with previous reported data with the conclusions drawn in Sec.~\ref{sec:conc}.

\section{Theoretical analysis}
\label{sec:theory}

\subsection{Model with blocking electrodes and the solution of the linearized PNP equations}

Let us consider an electrolyte layer bounded by parallel planar electrodes at a distance $\tilde x = \pm L$. 
Assuming the electrolyte contains completely dissociated negative and positive ions with the same mobility $\mu^- = \mu^+ = \mu$, diffusion coefficient $D^- = D^+=D$, valence $\hat{z}^- = \hat{z}^+=\hat{z}$, and initial concentration $c^-_0 = c^+_0 = c_0$, 
the PNP equations in 1-D case (see, e.g., \cite{Bazant:2004:PRE}) are 
\begin{eqnarray}
\label{eq:cp}
&& \dt{c^+} = -\dxtil{}\left( -D \dxtil{c^+} -\mu \hat{z}\, e\, c^+ \dxtil{\Phi} \right) , \\
\label{eq:cm}
&& \dt{c^-} = -\dxtil{}\left( -D \dxtil{c^-} + \mu \hat{z}\, e\, c^- \dxtil{\Phi} \right) , \\
\label{eq:phi}
&& -\varepsilon_0 \varepsilon \dxxtil{\Phi} =  \hat{z}\,e(c_+ - c_-) ,
\end{eqnarray}
where $c^+ = c^+(\tilde x, t)$, $c^-=c^-(\tilde x, t)$ are the concentrations of positive and negative ions, respectively, $\Phi = \Phi(\tilde x, t)$ is the electrostatic potential, $\varepsilon$ is the dielectric constant, $\varepsilon_0$ is the vacuum permittivity and $e$ is the elementary charge.

It is convenient to introduce dimensionless electrostatic potential $\psi$, relative difference in the concentrations between positive and negative ions  $\rho$, and dimensionless coordinate $x$
\begin{equation}
\label{eq:psi:rho:x}
\psi = \frac{\Phi}{U_T}, \quad \rho = \frac{c^+ - c^-}{2 \, c_0}, \quad x = \frac{\tilde x}{\lambda_D} ,
\end{equation} 
where the thermal voltage $U_T$ and the Debye length $\lambda_D$ are
\begin{equation}
\label{eq:lam}
 U_T = \frac{ k_B\, T}{\hat{z}\, e }, \quad 
\lambda_D = \sqrt{\frac{\varepsilon_0 \varepsilon\, k_B T}{2  c_0 \hat{z}^2 e^2}} ,
\end{equation}
with $k_B$  is the Boltzmann's constant and $T$ is the temperature.
Note, that at $T = 25^{\circ}$C for the ion valency  $\hat{z} = 1$ one has $U_T \approx 25$~mV, and 
typical values of the Debye length are within the range $\lambda_D = (1 - 100)$~nm for aqueous electrolytes.
At low applied voltages, $\Phi \ll U_T$ ($\psi \ll 1$), variations in the ion concentration are small, $\rho \ll 1$, and $(c^+ + c^-) \approx 2\, c_0$. 
Then, taking into account Eq.~(\ref{eq:psi:rho:x}), the Poisson equation (\ref{eq:phi}), and Einstein's relation $\mu = D / k_B\,T$, the equations (\ref{eq:cp}) and (\ref{eq:cm}) reduce to
\begin{equation}
\label{eq:rho:t}
\tau_q \dt{\rho} = \dxx{\rho} - \rho ,
\end{equation}
where $\tau_q$ is the charge relaxation time in the bulk given by
\begin{equation}
\label{eq:tauq}
 \tau_q = \frac{\lambda_D^2}{D}.
\end{equation}

For the applied ac voltage $U = U_0 \exp(i \omega t)$ ($\omega = 2 \pi f$ is an angular frequency) one has a linear response $(\rho, \psi) \sim \exp(i \omega t)$ with spatially dependent amplitudes of the relative difference in the concentrations between positive and negative ions $\rho(x)$ and the electrostatic potential $\psi(x)$ that can be found from
\begin{eqnarray}
\label{eq:rho}
&& \dxx{\rho} - k^2 \rho = 0 , \\
\label{eq:psi}
&& - \dxx {\psi} =  \rho ,
\end{eqnarray}
where
\begin{equation}
\label{eq:k}
 k = \sqrt{1+i\omega \tau_q}.
\end{equation} 

To model the effect of compact layers or polymer films at the electrodes we consider the case when the electrolyte and the electrodes are separated by two identical nonconducting dielectric layers of thickness $h$ and the dielectric constant $\varepsilon_p$.
We define an effective thickness of the dielectric layer $\lambda_p$ and the ratios of length scales $\delta$ and $\epsilon$ as following
\begin{equation}
\label{eq:lamp:del:eps}
 \lambda_p = h\, \frac{\varepsilon}{\varepsilon_p}, \quad \delta = \frac{\lambda_p}{\lambda_D},
 \quad \epsilon = \frac{\lambda_D}{L} .
\end{equation} 
For the blocking electrodes the boundary conditions can be formulated as follows \cite{Itskovich:1977:PSSA, Bazant:2004:PRE, Olesen:2010:PRE}
\begin{eqnarray}
\label{eq:bc:cur}
&& \left(\dx{\rho} + \dx{\psi}\right)_{x = \pm 1/\epsilon} = 0 ,  \\
\label{eq:bc:psi}
&& \mp v_0 -  \psi_0 = \pm \delta \dx{\psi}\Big|_{x = \pm1/\epsilon} ,
\end{eqnarray} 
where $v_0 = U_0  / U_T$ and $\psi_0$ are the values of the potential at the electrodes and at the boundary between the dielectric layer and the electrolyte, respectively.
The condition Eq.~(\ref{eq:bc:cur}) represents impermeability of the boundary for the ions and the absence of any physical or chemical processes at the boundary that may result to the ionic current.
The right hand side of equation (\ref{eq:bc:psi}) describes a voltage drop on the dielectric layer.

Taking into account the boundary conditions Eqs.~(\ref{eq:bc:cur}),  (\ref{eq:bc:psi}), the solutions of Eqs.~(\ref{eq:rho}), (\ref{eq:psi}) can be written as
\begin{eqnarray}
\label{eq:rho:sol}
  && \rho(x) = \rho_0 \frac{\sinh (k\, x)}{\sinh (k\, / \epsilon)} ,\\
\label{eq:psi:sol}
  && \psi(x) = -\frac{\psi_0}{k^2  z_e} \left(\epsilon\, x +  z_{w}\frac{\sinh (k\, x)}{\sinh (k\, / \epsilon)} \right) ,
 \end{eqnarray}
 where 
\begin{eqnarray}
 \label{eq:psi0}
 && \rho_0 = \psi_0 \, \frac{ z_{w}}{z_e }, \quad \psi_0 = v_0\, \frac{z_e}{z_e + \delta  \epsilon /(k^2-1)} , \quad \\
\label{eq:ze}
 && z_e =\frac{1}{k^2}  + \frac{z_{w}}{k^2}, \quad z_{w} = \epsilon\, \frac{\tanh(k\, / \epsilon)}{k\, (k^2-1)}.
\end{eqnarray}  
Here $z_e$ represents the dimensionless impedance of the binary electrolyte with blocking electrodes without dielectric layers that contains contributions from the bulk charge  ($1/k^2$) and diffuse layer ($z_w/k^2$) \cite{Coelho:1991:JNCS, Barbero:2005:LC}

Solutions (\ref{eq:rho:sol}), (\ref{eq:psi:sol}) describe the spatial distribution of the electrical potential and the ionic charge density across the layer (the total charge density is proportional to $\rho$). 
According to Eq.~(\ref{eq:rho:sol}) the charge density has a maximum at the boundary between the dielectric layer and the electrolyte,  $|\rho(x=\pm1/\epsilon)| =\rho_0$, and decays exponentially with the distance from the boundary with the characteristic decay length $\lambda =  \Re{[\lambda_D / \sqrt{1+i \omega \tau_q}]}$ in physical units. 
The decay length $\lambda$ can be considered as a diffuse layer thickness, and is negligible at high frequencies, $\omega \tau_q \gg 1$, grows as frequency decreases and approaches the Debye length $\lambda_D$ in the limit of a dc voltage, $\omega \tau_q \to 0$. 
Note, that apart from simple frequency dependence the diffuse layer thickness is solely determined by the properties of the charge carriers -- diffusion coefficient and concentration.

The diffuse charge, $\rho_0$, in addition, is influenced by the properties of the dielectric layer.
Equations (\ref{eq:psi0}) show that $\rho_0$ is determined by the contribution of the diffuse layer $z_w$ to the total impedance $z_e$ and depends on the surface potential $\psi_0$ which is different from the electrode potential $v_0$ when the effective thickness of the dielectric layer is nonzero, $\delta \ne 0$.
It can be seen by considering the limiting cases of high and low frequencies.
Taking into account (\ref{eq:k}),  (\ref{eq:ze}), from (\ref{eq:psi0}) it follows that $\psi_0 \to  v_0 / (1 + \epsilon\delta)$ at high frequencies and approaches the value $\psi_0 \to v_0 / (1 + \delta)$ in the limit of a dc voltage, $\omega \tau_q \to 0$. 
Hence, a voltage drop on the dielectric layer $|v_0 - \psi_0|$ grows with decreasing frequency and its value depends on $\delta$.
Substituting  (\ref{eq:k}),  (\ref{eq:ze}) into (\ref{eq:psi0}), one finds $\rho_0 \to 0$ at $\omega \tau_q \to \infty$, and $\rho_0 \to v_0 /(1+ \delta)$ at $\omega \tau_q \to 0$.
Thus, the voltage drop across the dielectric layer reduces the diffuse charge and, hence, decreases the electrode polarization.
The ratio $\delta = \lambda_p / \lambda_D$ does not depend on the distance between electrodes $L$ (compare with \cite{Sorensen:1995:JCSFT}) and in the case of $\lambda_p \ge \lambda_D$  the influence of the dielectric layer on the impedance behavior can be essential, especially, at low frequencies.

\subsection{Impedance in the model with blocking electrodes}

An impedance is defined as a ratio of the applied voltage to the electrical current arising in the system.
In the case of blocking electrodes a current density in the external circuit is equal to a displacement current density $J_D = -\varepsilon_0 \varepsilon( \partial^2 \Phi / \partial t\, \partial \tilde x)$ at the electrode surface.
Denoting the density of the Nernst's diffusion-limited current $J_N = \hat{z} e c_0 D / L$ \cite{Bonnefont:2001:JEAC}, the normalized displacement current density can be written as
 \begin{equation}
 \label{eq:jd}
   j_D = \frac{J_D}{J_N} = -i \omega \frac{\tau_q}{\epsilon} \dx{\psi}\Big|_{x=-1/\epsilon}. 
 \end{equation} 
Then the dimensionless impedance has a form
 \begin{equation}
 \label{eq:def:z}
 z \equiv \frac{Z}{R} = \frac{v_0}{j_D} ,
 \end{equation}
 where $Z = U_0/(S\,J_D)$ is the impedance of a plane sample with the area $S$ and the resistance 
 \begin{equation}
 \label{eq:R}
 R \equiv \frac{U_T}{S\, J_N}=  \frac{k_B T L}{c_0 (z\, e)^2 D S}.
 \end{equation} 
Introducing the electrolyte capacitance $C = \varepsilon_0 \varepsilon S / (2 L)$, the well-known expression for the charge relaxation time in the bulk (RC-time) can be recovered  $\tau_q = R\, C = \lambda_D^2 / D$.

Calculating the displacement current in Eq.~(\ref{eq:jd}) from the solution $\psi(x)$ [Eqs.~(\ref{eq:psi:sol}), (\ref{eq:psi0})] and substituting $j_D$ into the definition of the dimensionless impedance Eq.~(\ref{eq:def:z}), we arrive at the final expression
\begin{equation}
 \label{eq:z}
  z = z_e + \frac{\delta\, \epsilon}{k^2-1} .
\end{equation} 
Here the last term represents a capacitive contribution of the dielectric layer.
According to Eqs.~(\ref{eq:k}),  (\ref{eq:ze}) this expression can be written as an explicit function of frequency
\begin{equation}
\label{eq:imp:exact}
 z = \frac{1}{1 + i \omega \tau_q} +  \frac{1}{i \omega \tau_c}  \frac{\tanh[\sqrt{1 + i \omega \tau_q} (\tau_c / \tau_q) ]}{(1 + i \omega \tau_q)^{3/2}} +  \frac{1}{i \omega \tau_p} ,
\end{equation} 
where
\begin{equation}
\label{eq:taucp}
 \tau_c \equiv \frac{\tau_q}{\epsilon} = \tau_q \frac{L}{\lambda_D}, \quad \tau_p \equiv \frac{\tau_q}{\epsilon\, \delta} =\tau_q \frac{L}{\lambda_p} .
\end{equation} 
The first term in (\ref{eq:imp:exact}) describes the bulk charge contribution to the total impedance and is a well-known expression for the dimensionless impedance of a parallel RC-circuit with characteristic time $\tau_q = R C$.
The second and the third terms represent the contributions of the two diffuse layers and the two dielectric layers, respectively.
Introducing capacitances $C_d = \varepsilon_0 \varepsilon_p S / \lambda_D$ and $C_p = \varepsilon_0 \varepsilon_p S / h$, and using  Eqs.~(\ref{eq:R}), (\ref{eq:lam}), one finds $\tau_c = R\, C_d / 2$ and $\tau_p = R\, C_p / 2$ representing the charging times of two diffuse and two dielectric layers, respectively.
Note, that using the results of Refs~\cite{Trukhan:1963:SPSS, Sorensen:1995:JCSFT} and setting the same mobility of positive and negative ions the expression for the impedance Eq.~(\ref{eq:imp:exact}) can be recovered.

Thus, the impedance of the binary electrolyte bounded by the electrodes with the dielectric layers includes three contributions with different frequency dependence.
To understand a role of each contribution to the overall behavior of the impedance analysis of limiting cases and approximate expansions can be useful. 

\subsection{Approximate expression for the impedance}
\label{sec:theory:appr:imp}

To simplify the analysis of the frequency dependence of the impedance (\ref{eq:imp:exact}), we consider the limit of thin diffuse layer, $\epsilon = \lambda_D / L \ll 1$, which is a typical case for electrolytes with the Debye length $\lambda_D = (1 - 100)$~nm and the sample thickness $L > 1$~$\mu$m. 
The expression (\ref{eq:imp:exact}) can be rewritten as a sum of the bulk contribution $z_v$ (the first term) and the surface one $z_s$ (the sum of the second and the third terms).
In the limiting case $\epsilon \ll 1$ one has $\tanh[\sqrt{1 + i \omega \tau_q} (\tau_c / \tau_q)] \approx 1$ and $z_v$ and $z_s$ can be expanded in a series of $\omega$. 
Taking into account Eq.~(\ref{eq:taucp}) the expansion for the high frequency range $\omega \tau_q \gg 1$ will take the form
\begin{eqnarray}
\label{eq:imp:high:v}
  && z_v =  \frac{1}{(\omega \tau_q)^2} +  \frac{1}{i \omega \tau_q}  + O(\omega^{-3}) ,\\
\label{eq:imp:high:s}
  && z_s =  \frac{ \delta\, \epsilon}{i \omega \tau_q} 
  + \frac{\epsilon}{(i \omega \tau_q)^{5/2}} + O(\omega^{-3}) ,
\end{eqnarray}
and for the low frequencies, $\omega \tau_q \ll 1$,
\begin{eqnarray}
\label{eq:imp:low:v}
 && z_v =  1  - i \omega \tau_q  + O(\omega^{2}) , \\
 \label{eq:imp:low:s}
 && z_s =  \epsilon \left(- \frac{3}{2}+  \frac{15}{8} i \omega \tau_q\right)   + \epsilon  \frac{ 1 + \delta}{i \omega \tau_q} 
  + O(\omega^{2}).
\end{eqnarray}
From Eqs.~(\ref{eq:imp:high:v}), (\ref{eq:imp:high:s}) it follows that the high frequency behavior of the impedance in the leading order in $\epsilon$ is only determined by the bulk charge dynamics with the characteristic time $\tau_q$ represented by the first term in the expression (\ref{eq:imp:exact}).
The expansion (\ref{eq:imp:high:v}) shows that the real and imaginary parts of the impedance depend on frequency as $(\omega\tau_q)^{-2}$  and $(\omega\tau_q)^{-1}$, respectively.
According to Eqs.~(\ref{eq:imp:low:v}), (\ref{eq:imp:low:s}) at low frequencies the real part of the impedance is constant ($z  = 1 + O(\epsilon)$) and the frequency dependence of the imaginary part for $\omega \to 0$ will be determined by the surface contribution $z_s = 1/(i \omega \tau_s)$  with the characteristic time
\begin{equation}
\label{eq:taus}
 \tau_s \equiv \frac{\tau_q}{\epsilon\,(1+\delta)} = \tau_q \frac{L}{\lambda_s} ,
\end{equation} 
where
\begin{equation}
\label{eq:lams}
  \lambda_s =  \lambda_D + \lambda_p .
\end{equation}
The expressions  (\ref{eq:taus}),  (\ref{eq:taucp}) show that $\tau_s$ is the combination of the charging times of the diffuse and dielectric layers
\begin{equation}
\label{eq:taus:cp}
\frac{1}{\tau_s} = \frac{1}{\tau_c} + \frac{1}{\tau_p} .
\end{equation}
Compare the imaginary parts of (\ref{eq:imp:high:s}), (\ref{eq:imp:low:s}), it can be seen, that the surface contribution to the impedance dominates over the bulk one in the low frequency range  $\omega < \omega_{qs}$, where
\begin{equation}
\label{eq:oms}
 \omega_{qs} = \sqrt{\frac{1}{\tau_q \tau_s}} =\frac{1}{\tau_q} \sqrt{ \frac{\lambda_D + \lambda_p}{L}} .
\end{equation} 
Thus, the impedance behavior in the high frequency range $\omega \gg \omega_{qs}$ is solely determined by the bulk contribution $z_v = 1/(1 + i \omega \tau_q)$ and $z_s$ is negligible.
On the contrary, at low frequencies $\omega \ll \omega_{qs}$ the surface contribution $z_s = 1/(i \omega \tau_s)$ is predominant.
An approximate expression for the impedance can be written as a sum of these contributions
\begin{equation}
\label{eq:imp:sim}
z = \frac{1}{1 + i \omega \tau_q} +  \frac{1}{i \omega \tau_s} .
\end{equation}
This expression can be interpreted as the dimensionless impedance of the parallel RC-circuit connected in series with capacitor, whose capacitance is $C_s = 2\, \tau_s / R = \varepsilon_0 \varepsilon S / \lambda_s$.
Multiplying Eq.~(\ref{eq:taus:cp}) by $R$, it follows that the capacitance $C_s = (1/C_d + 1/C_p)^{-1}$ is represented by the diffuse and dielectric layer capacitances connected in series (see, e.g., \cite{Bazant:2004:PRE, Olesen:2010:PRE}).
However, it is important to note, that $\lambda_s$ is the sum of the Debay length and the effective thickness of the polymer layer, as it follows from Eq.~(\ref{eq:lams}).
Hence, the diffuse and the dielectric layers act as a single capacitive layer with the thickness $\lambda_s$, that determines the behavior of the impedance in the low frequency range.

Thus, in the most common practical cases of the thin diffuse layer, $\lambda_D \ll L$, the frequency dependence of the exact expression for the impedance  (\ref{eq:imp:exact})  in the leading order in $\epsilon$ is equivalent to the frequency dependence of the approximate expression  (\ref{eq:imp:sim}).
Apart from the simplicity of the expression  (\ref{eq:imp:sim}), it shows, that the impedance behavior is determined by the two different mechanisms -- dynamics of the bulk charge and the simultaneous charging of the diffuse and the dielectric layers with the corresponding  characteristic times $\tau_q$ and $\tau_s$.
Each of the mechanisms dominates in distinct frequency domains separated by the frequency $\omega_{qs}$, in the vicinity of which the contributions are competing, that can be discovered in the impedance spectra peculiarities.

\subsection{Analysis of the impedance frequency dependence}
\label{sec:theory:freq_analysis}

The approximate expression for the impedance (\ref{eq:imp:sim}) makes it easy to analyze frequency peculiarities of the impedance spectra.
Representing the dimensionless impedance (\ref{eq:imp:sim}) in the form  $z = z' - i z''$, the real part can be written as
\begin{equation}
 z' =\frac{1}{1 + \omega^2 \tau_q^2} .
\end{equation}  
This is a monotonic function of the frequency with $z' \sim (\omega\tau_q)^{-2}$ for $\omega \tau_q\gg 1$.
In the limiting cases one has $z' \to 0$ for $\omega \to \infty$ and $z' \to 1$ for $\omega \to 0$.
The real part of the impedance reaches a plateau below the frequency
\begin{equation}
\label{eq:omq}
 \omega_q = \frac{1}{\tau_q} ,
\end{equation} 
at which it has the value $z'(\omega = \omega_q) =1/2$.

The imaginary part 
\begin{equation}
\label{eq:imp:simp:im}
 z'' = \frac{1}{\omega \tau_s} +  \frac{\omega \tau_q}{1 +(\omega \tau_q)^2}
\end{equation} 
may have two local extrema at frequencies
\begin{equation}
\label{eq:ext:im}
  \omega_{1,2} = \frac{1}{\tau_q} \sqrt{\frac{1}{2}\frac{ \left(1  \pm \sqrt{1 - 8\, \tau_q / \tau_s} \right) - 2\, \tau_q / \tau_s }{1 + \tau_q / \tau_s}}.
\end{equation}  
For $\tau_q/\tau_s = \epsilon(1 + \delta) \ll 1$ the maximum and the minimum of $z''$ are located at
\begin{equation}
\label{eq:om:im}
 \omega_{max}  = \sqrt{\frac{1}{\tau_q^2} - \frac{4}{\tau_q \tau_s}} \approx \omega_q, \quad \omega_{min} = \omega_{qs}
\end{equation} 
with corresponding values
\begin{equation}
  z''_{max} \approx  \frac{1}{2} + \frac{\tau_q}{\tau_s}, \quad z''_{min} \approx \sqrt{\frac{\tau_q}{\tau_s}} .
\end{equation} 
As follows from  (\ref{eq:ext:im}), the local extrema in the imaginary part of the impedance appear only if the condition $\tau_s > 8\, \tau_q$ is satisfied.
Otherwise, the imaginary part of the impedance (\ref{eq:imp:simp:im}) will be a monotonic function of frequency. 
Increase in the concentration of the charge carriers (the decrease of $\lambda_D$ and, hence, $\tau_q$) shifts the extrema positions of  $z''$ to the high frequency range and makes deeper a local minimum, such that  ($z''_{max} -z''_{min}) \to 1/2$ for $\tau_s \gg \tau_q$.
Increase of the distance between electrodes $L$ or decrease of the effective thickness of the dielectric layer $\lambda_p$ move the minimum  $z''$ to the lower frequencies without changing the maximum position, as it follows from (\ref{eq:taus}), (\ref{eq:oms}), (\ref{eq:omq}) and (\ref{eq:om:im}).

The impedance argument
\begin{equation}
\label{eq:imp:sim:arg}
  \arg{z} = - \arctan \left[ \frac{1}{\omega \tau_s} +  \left(1 + \frac{\tau_q}{\tau_s} \right)  \omega \tau_q \right] 
 \end{equation}
has a local maximum at the frequency
\begin{equation}
\label{eq:arg:max:om}
 \omega_{m} = \sqrt{\frac{1}{\tau_q \tau_s(1 + \tau_q/\tau_s)}},
\end{equation} 
where its value
\begin{equation}
\label{eq:arg:max}
  a_{m} \equiv -\tan \left( \arg{z} \right)_{max} = 2 \sqrt{\frac{\tau_q}{\tau_s} \left(1 + \frac{\tau_q}{\tau_s} \right)}.
\end{equation} 
This allows to express the characteristic times in terms of $\omega_m$ and $a_m$
\begin{equation}
  \tau_q = \frac{a_{m}}{\omega_{m} (\sqrt{1+a_{m}^2} - 1)}, \quad \tau_s = \frac{2}{a_{m} \omega_{m}}.
\end{equation} 
For $\tau_s \gg \tau_q $ the frequency $\omega_{m} \approx \omega_{qs}$ and the value $a_{m} \approx 2 \sqrt{\tau_q / \tau_s}$, hence, the characteristic times can be found from the simple relations 
\begin{equation}
\label{eq:tqts:arg}
 \tau_q = \frac{a_{m}}{2 \omega_{qs}}, \quad \tau_s = \frac{2}{a_{m} \omega_{qs}} .
\end{equation} 
As it can be seen, in the case of $\tau_s \gg  \tau_q$ the frequency of the $\arg{z}$ maximum  coincides with the frequency of $z''$ minimum and is defined by $\omega_{qs}$ [Eq.~(\ref{eq:oms})], at which the bulk and the surface contributions are nearly equal.
The maximum value $\arg{z}_{max} \approx \arctan(-2 \sqrt{\tau_q/\tau_s})$ will be larger for the larger values of the concentration of the charge carriers and the distance between electrodes, and smaller for the smaller effective thickness of the dielectric layer [see Eq.~(\ref{eq:taus})].

Thus, in the thin diffuse layer limit, the positions and extrema values of the imaginary part and the impedance argument are uniquely determined by characteristic times $\tau_q$ and $\tau_s$.
The charge relaxation time $\tau_q$ determines the frequency of the imaginary part maximum. 
The charging time of the diffuse and dielectric layers,  $\tau_s$, defines the positions of the minimum of $z''$ and the maximum of $-\tan(\arg{z})$, which coincide for $\tau_s \gg \tau_q$.
Note, that according Eq.~(\ref{eq:taus}), (\ref{eq:lams})  the dielectric layer may strongly influence these positions in the case of $\lambda_p \ge \lambda_D$.

\subsection{Calculations of ion parameters}
\label{sec:theory:ionpars}
The diffusion coefficient $D$ and the concentration of the charge carriers  $c_0$ can be found from the positions and values of the extrema of the impedance argument and imaginary part $z''$.
However, accurate localization of the extrema in the experimental data requires a high frequency resolution of the impedance spectra, that may result in time-consuming measurements.
Another approach is a fitting of experimental impedance spectra with an appropriate model \cite{Sorensen:1995:JCSFT, Klein:2006:JCP,Alexe-Ionescu:2009:JAP, Paula:2012:PRE}.
Consider the exact, Eq.~(\ref{eq:imp:exact}), and the approximate, Eq.~(\ref{eq:imp:sim}), expressions for the impedance written in physical units
\begin{eqnarray}
\label{eq:imp:exact:dim}
Z_{ex} =  && R \left(  \frac{1}{1 + i \omega \tau_q} +    \frac{1}{i \omega \tau_p}  \right. 
 \nonumber   \\ 
 &&  \left. +  \frac{1}{i \omega \tau_c} \frac{\tanh[\sqrt{1 + i \omega \tau_q} (\tau_c / \tau_q)]}{(1 + i \omega \tau_q)^{3/2}}  \right), \\
 \label{eq:imp:simp:dim}
Z_{ap} =  &&  R \left( \frac{1}{1 + i \omega \tau_q} +  \frac{1}{i \omega \tau_s} \right) .
\end{eqnarray} 
The expression (\ref{eq:imp:exact:dim}) contains four parameters $R$, $\tau_q$, $\tau_c$ and $\tau_p$, which can be determined by the nonlinear least-squares fitting of experimental spectra.
Using the fitted parameters, the Debye length and the effective thickness of the dielectric layer can be found [see Eq.~(\ref{eq:taucp})]
\begin{equation}
 \lambda_D = \frac{\tau_q}{\tau_c} L, \quad \lambda_p = \frac{\tau_q}{\tau_p} L .
\end{equation}

In the case of $\lambda_D \ll L$ the frequency dependence of the impedance is described by the approximate expression (\ref{eq:imp:simp:dim}). 
Here the fitting provide only three parameters, $R$, $\tau_q$, and $\tau_s$, and if the effective thickness of the dielectric layer $\lambda_p$ is known, the Debye length can be calculated from 
\begin{equation}
\label{eq:lamdp}
 \lambda_D = \frac{\tau_q}{\tau_s} L - \lambda_p ,
\end{equation}
as it follows from  Eq.~(\ref{eq:taus}), (\ref{eq:lams}).

Finally, using $\lambda_D$ and $\tau_q$, the values of the diffusion coefficient $D$, the concentration of the ions $c_0$, and the hydrodynamic radius of ions $R_g$ can be found from 
\begin{equation}
\label{eq:DR}
 D = \frac{\lambda_D^2}{\tau_q},  \quad   c_0 =  \frac{\varepsilon_0 \varepsilon\, k_B T}{2\, \lambda_D^2 (\hat z\, e)^2},\quad R_g = \frac{k_B T }{6 \pi \eta D} ,
\end{equation} 
according to the definitions (\ref{eq:lam}), (\ref{eq:tauq}), and using the well-known Stocks formula \cite{Landau:1987}; here $\eta$ is a medium viscosity.
Unlike similar relationships obtained in Refs.~\cite{Klein:2006:JCP, Munar:2011:JNCS}, in this approach it is not necessary to know the value of $R$ (or dc-conductivity) to determine $c_0$ and $D$.
However, for known $R$, $\tau_q$, and the capacitance of the empty cell $C_0 = \varepsilon_0 S/ 2 L$, the conductivity and the dielectric constant of the electrolyte can be easily calculated from
\begin{equation}
\label{eq:sigmaepsilon}
 \sigma = \frac{\varepsilon_0}{R C_0}, \quad \varepsilon = \frac{\tau_q}{ R C_0}.
\end{equation} 

\section{Experiment}
\label{sec:exp}

\subsection{Cells preparation and impedance measurements}
Impedance spectra of the liquid crystal 5CB were measured in cells assembled of two plane parallel glass substrates with a transparent conductive layer coated by a thin polymer film.
The substrate conductive layers of the indium-tin oxide (ITO) were chemically etched to make square electrode areas ($S =10 \times 10$~mm$^2$).
The polyimide solution JALS-204 (JSR, Japan) was spin-coated on the top of electrodes according to a procedure described by the manufacturer.
The thickness of the polymer film measured by a interferometer MII-4 (LOMO, Russia) and an atomic-force microscope Agilent 5500 AFM (Agilent, USA) was $h = (30 \pm 2)$ nm for all substrates.

A gap between the substrates was fixed by mylar spacers or thin layer of the UV-glue around the cell in the case of thin samples (below 10~$\mu$m).
Gap thickness $d$ was measured in empty cells at several points inside of the electrode area using a spectrometer; in all cells the thickness heterogeneity was less than $0.5$~$\mu$m/cm. 
The empty cell capacitance $C_0 = \varepsilon_0 S / d$ was measured by a RLC meter.
The values for the thickness calculated from $C_0$ and measured by the spectral method were in agreement within the accuracy of both methods.

The nematic liquid crystal 5CB (TCI, Europe) was filled into the cells in an isotropic phase at $T=45^{\circ}$C, slowly cooled down and kept at room temperature for 2 hours before impedance spectra measurements.
Observations in a polarizing microscope demonstrated homogeneous homeotropic orientation (the director is oriented everywhere perpendicular to the substrates). 
Thus the dielectric permittivity of the 5CB was spatially homogeneous and equal to $\varepsilon_{\parallel}$ (parallel to the director).

All measurements were performed at the temperature $T = 25^{\circ}$C.
The impedance spectra were measured by a potentiostat AutoLab (Metrohm Autolab B.V., Netherlands) with a FRA32 module by a two-electrode cell setup applying an ac voltage with the amplitude 25~mV  and frequency in the range ($10^{-4} - 10^5$)~Hz.
Note, that the dielectric anisotropy $\varepsilon_a =\varepsilon_{\parallel} - \varepsilon_{\perp}$ of the liquid crystal 5CB is positive and the applied voltage stabilizes the initial homeotropic orientation of the LC layer.
Hence, all obtained parameters relate to the parallel components of corresponding tensors; in the following the indices $\parallel$ in the notation of physical quantities will be omitted.

\subsection{Impedance spectra}
\begin{figure}
\centering
\includegraphics[width=\figsize]{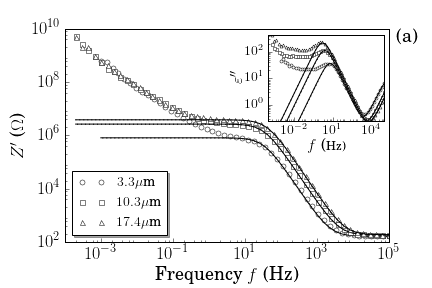}
\includegraphics[width=\figsize]{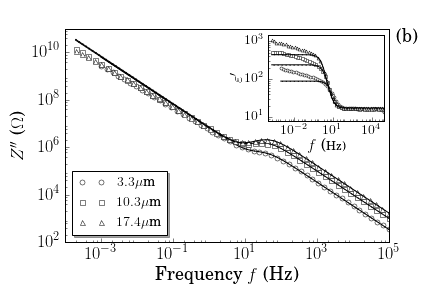}
\includegraphics[width=\figsize]{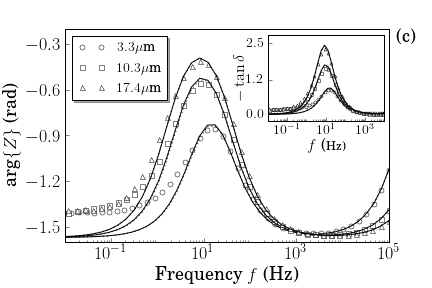}
\includegraphics[width=\figsize]{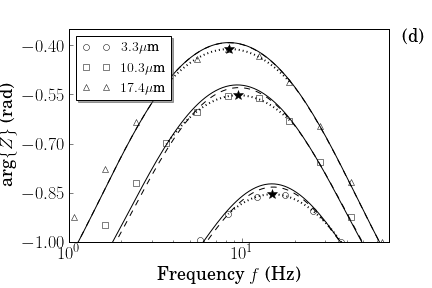}
\caption{
Impedance spectra of the cells with different thickness (dielectric spectra in insets) (a) -- (c). The symbols indicate experimental data, the solid and dashed lines denotes the fitted curves $Z_{ex}$ by Eq.~(\ref{eq:imp:exact:dim}) and $Z_{ap}$ by Eq.~(\ref{eq:imp:simp:dim}), respectively. 
(d) Argument of the impedance $\arg{Z}$ in the region of its maximum: polynomial interpolation shown by the dotted lines with the maximums marked by the stars.}
 \label{fig:imp-fresh}
\end{figure}

Figure~\ref{fig:imp-fresh} shows the real $Z'$ and imaginary  $Z''$ parts of the impedance $Z = Z' - i Z''$,  and its argument $\arg{Z}$ for the cells of different thickness.
The corresponding frequency dependence of the real and imaginary parts of the complex dielectric constant $\varepsilon' - i \varepsilon'' =1/(i \omega Z C_0)$ and the dielectric loss tangent $\tan \delta = \varepsilon''/\varepsilon'$ are given in insets of Fig.~\ref{fig:imp-fresh} for comparison with previous reported data on 5CB \cite{Murakami:1997:JJAP,  Sawada:1999:JJAP, Naemura:2003:MCLC}.
It can be seen, that the impedance (dielectric) spectra of the liquid crystal 5CB are typical for materials with ionic conductivity \cite{Klein:2006:JCP, Sangoro:2008:PRE, Serghei:2009:PRB,Munar:2011:JNCS, Wang:2013:PRE}.

At the high frequencies,  ($10^{4}-10^{5}$)~Hz, the real part of the impedance, $Z'$, has a plateau [Fig.~\ref{fig:imp-fresh}(a)] with the level defined by the resistance of cell contacts with an external circuit ($R_c \approx (100-200)$~$\Omega$).
With decreasing frequency $\re{Z}$ increases as $f^{-2}$ and goes to another plateau corresponding to the resistance of the liquid crystal layer ($R \approx (1-10)$~M$\Omega$ depending on the LC layer thickness).
The imaginary part of the impedance, $Z''$, monotonically increasing as $f^{-1}$ up to  $f \approx 100$~Hz with decreasing frequency (Fig.~\ref{fig:imp-fresh}b).
Below $f \approx 100$~Hz there are two local extrema for the cells with the thickness $d =17.4$~$\mu$m and $d = 10.3$~$\mu$m; they are missing in the spectrum of the thinnest cell with $d = 3.3$~$\mu$m.
The argument $\arg{Z}$  has a pronounced maximum around $10$~Hz  [Figs.~\ref{fig:imp-fresh}(c)].
For the cells with larger thickness the maximum value of the argument is larger and the position is shifted to the lower frequencies.

In the frequency range $f \gtrsim 1$~Hz the impedance spectra in all cells are qualitatively well described by Eq.~(\ref{eq:imp:exact:dim}).
However, for $f  \lesssim 1$~Hz the impedance behavior deviate from that given by Eq.~(\ref{eq:imp:exact:dim}). 
In particular, $Z'$ increases and $Z''$ deviates from the $f^{-1}$-dependence with decreasing frequency, that is also reflected in a low frequency behavior of the real part of the dielectric constant $\varepsilon'$, which is increasing instead of being a constant in the limit $f \to 0$.
To explain such behavior of the impedance spectra, different physical mechanisms have been proposed, ranging from a fractal nature of electrode surfaces \cite{Pajkossy:1994:JEAC} or double layers \cite{Feldman:1998:PRE} to the adsorption processes with an anomalous diffusion \cite{Paula:2012:PRE}.
Our PNP model does not take into account any of those contributions, consequently, experimental data will be analyzed only in the frequency range  $f \gtrsim 1$~Hz.

\subsection{Fitting of the impedance spectra}
The absence of the extrema in the imaginary part of the impedance, $Z''$, for the cell with the thickness  $d=3.3$~$\mu$m indicates possible violation of the condition $\lambda_D \ll L$ used in the derivation of the approximate expression for the impedance. 
Therefore, the spectra have been fitted by means of the exact expression Eq.~(\ref{eq:imp:exact:dim}) and the approximate one Eq.~(\ref{eq:imp:simp:dim}) for comparison.
Initial values of the model parameters $\tau_q$,  $\tau_s$, and $R$ were determined from the position and the maximum value of the argument $\arg{Z}$  using  Eq.~(\ref{eq:tqts:arg}) and the plateau level of the real part $Z'$.
Initial values of $\tau_c$ and $\tau_p$ required for the exact expression were calculated from  Eqs.~(\ref{eq:taucp}), (\ref{eq:taus:cp}) for the parameters of the polymer layer
\begin{equation}
\label{eq:polpar} 
  h = 30 \mbox{ nm}, \quad \varepsilon_p = 3.5 .
\end{equation}

To determine a frequency range, where the experimental spectra are best described by the proposed model, first, the position of the argument maximum $f_{max}$ has been localized by a polynomial interpolation (starts symbols on Fig.~\ref{fig:imp-fresh}d).
Then the spectra were fitted using the expressions (\ref{eq:imp:exact:dim}) and (\ref{eq:imp:simp:dim}) in the frequency range $f > f_0$ with $f_0$ varying from 0.1~Hz to 10~Hz and the frequency $f^{fit}_{max}$ corresponding to the argument maximum was calculated for each $f_0$ chosen.
A frequency range with the value of $f_{max}^{fit}$ closest to the interpolated experimental value $f_{max}$ was accepted as the best choice and the model parameters obtained by fitting over this frequency range were used for further calculations.

An attempt to fit all four parameters  ($R$, $\tau_q$, $\tau_c$, and $\tau_p$), using the exact expression  (\ref{eq:imp:exact:dim}),  revealed the linear dependence between parameters $\tau_c$ and $\tau_p$, which was manifested in large scatter of their values for different cells.
To avoid this, we decreased the number of varied parameters to three ($R$, $\tau_q$, and $\tau_c$) substituting $\tau_p = R\, C_p / 2$ into  Eq.~(\ref{eq:imp:exact:dim}), where $C_p =C_0 \varepsilon_p d/h$ was calculated for each cell using the parameters of the polymer layer (\ref{eq:polpar}).

Fitted curves, shown in Fig.~\ref{fig:imp-fresh} by solid lines, demonstrate a good agreement with experimental data for all cells in the frequency range $f \gtrsim 1$~Hz.
The curves, corresponding to the exact and the approximate expressions, practically coincide; small deviations are only observed in the vicinity of the maximum of  $\arg{Z}$ for the thin cells with $d=3.3$~$\mu$m and $d=10.3$~$\mu$m (see Fig,~\ref{fig:imp-fresh}d).

Table~\ref{tab:fitpars-fresh} contains the fitted parameters $R$, $\tau_q$, $\tau_c$, obtained by use of the exact (\ref{eq:imp:exact:dim}) and the approximate (\ref{eq:imp:simp:dim}) expressions for the impedance, and the values of the dielectric constant $\varepsilon$ and conductivity $\sigma$ of the liquid crystal calculated from Eq.~(\ref{eq:sigmaepsilon}).
For each LC cell characteristic time $\tau_s$ in the upper part and $\tau_c$ in the low part of the row were found using  Eq.~(\ref{eq:taus:cp}) with  $\tau_p$  calculated from Eq.~(\ref{eq:taucp}).
Note, that the conductivity $\sigma$ is noticeably scattered (from 33.1~nS/m to  40.6~nS/m) for different cells as well as obtained from the exact and the approximate expressions for the impedance.
This can result from a weak variations in the LC cell preparation, e.g., quality of the substrate cleaning or amount of UV-glue contacted with the liquid crystal at the cell edges, that influence the ion concentration \cite{Mada:1996:JJAP}.
Obtained values of the dielectric constant $\varepsilon$ tend to decrease with increasing cell thickness, nevertheless, they agree quite well with the previously reported data on $\varepsilon_{\parallel}$ for 5CB ranging from $17.6$ to $20.2$~\cite{Bogi:2001:LC}.
Small difference between the values of $\varepsilon$ obtained by using the exact and the approximate expressions for the impedance decreases with increasing the cell thickness.
For the thickest cell with $d = 17.4$~$\mu$m both expressions give almost the same value $\varepsilon \approx 17$.

To conclude, the exact expression for the impedance fits the experimental spectra for thin cells better.
Nearly equal values of $\varepsilon$, $\sigma$, obtained for the cell with the thickness $d = 17.4$~$\mu$m confirm that Eq.~(\ref{eq:imp:simp:dim}) is indeed a good approximation for the exact expression for the impedance (\ref{eq:imp:exact:dim}), if the condition $\lambda_D \ll L$ is fulfilled.
Note, that the decrease in the dielectric constant  with increasing cell thickness is probably due to an incompleteness of the considered PNP model, which is applicable only in a limited frequency range.

\begin{table}
\caption{Fitted parameters for different LC cells ($R$, $\tau_q$, $\tau_c$, $\tau_p$) and calculated values ($\varepsilon$, $\sigma$). For each LC cell, the parameters obtained from $Z_{ex}$ (upper row) and $Z_{ap}$ (low row) are listed.}
\begin{center}
\begin{ruledtabular}
\begin{tabular}{*7c}
$d$,~$\mu$m & $R$, M$\Omega$ & $\tau_q$, ms & $\tau_c$, ms & $\tau_s$, ms & $\varepsilon$ & $\sigma$, nC/m\\\hline
\multirow{ 2}{*}{$ 3.3$}  & 0.96 &   4.95 &  42.7 &  23.3 &  18.5 &  33.1 \\
&   0.81 &   4.70 &  37.1 &  20.0 &  20.9 &  39.4 \\\hline
\multirow{ 2}{*}{$10.3$} & 2.81 &   4.60 &  99.6 &  61.2 &  17.4 &  33.4 \\
& 2.65 &   4.55 &  98.5 &  59.4 &  18.2 &  35.4 \\\hline
\multirow{ 2}{*}{$ 17.4$} & 3.98 &   3.83 & 163 &  95.6 &  16.9 &  39.3 \\
& 3.84 &   3.78 & 158 &  92.1 &  17.4 &  40.6
\end{tabular}
\end{ruledtabular}
\end{center}
\label{tab:fitpars-fresh}
\end{table}

Using the results from Table~\ref{tab:fitpars-fresh} the ion parameters are calculated from Eqs.~(\ref{eq:lamdp}), (\ref{eq:DR}) (see Table~\ref{tab:ionpars-fresh}).
The effective thickness of the polymer layer for all three cells was found around $\lambda_p = h \varepsilon/\varepsilon_p =150$~nm; small variations ($<10$~nm) came out due to the variation of $\varepsilon$ for the different cells.
To calculate the hydrodynamic radius of the ions, the average effective viscosity of the Stokes drag in the liquid crystal 5CB $\eta = 0.06$ Pa$\cdot$s was taken \cite{Stark:2001:PREP}.
\begin{table}
\caption{The calculated  ion parameters for different LC cells.}
 \begin{center}
 \begin{ruledtabular}
 \begin{tabular}{*6c}
$d$, $\mu$m  & $\lambda_D$, nm & $c_0$, $\mu$m$^{-3}$ & $D$, $\mu$m$^2$/s & $R_g$, nm & $D^*$, $\mu$m$^2$/s\\\hline
$ 3.3$  & 209  & 339 &   9.3 &    0.43 &  32.1 \\\hline
$ 10.3$ & 237 & 230 &  12.4 &   0.32 &  34.0 \\\hline
$ 17.4$ & 208 & 284 &  11.5 &   0.35 &  33.7 \\
 \end{tabular}
 \end{ruledtabular}
 \end{center}
 \label{tab:ionpars-fresh}
 \end{table}

The results in Table~\ref{tab:ionpars-fresh} show that the values of the Debay length are about $\lambda_D \approx 200$~nm and vary within $30$~nm for the cells of different thickness, that is respectively reflected in the variations of the concentration of the ions $c_0$ and the diffusion coefficient $D$.
The values of $D$ are nearly the same for all cells and $D \approx 12$~$\mu$m$^2$/s can be taken as a typical value of the diffusion coefficient of ions in the liquid crystal 5CB.
This value of the diffusion coefficient corresponds to the hydrodynamic radius of the ion $R_g \approx 0.3$~nm, that is close to a typical size of free inorganic ions \cite{Israelachvili:2011}.
For comparison, the last column in Table~\ref{tab:ionpars-fresh} contains the values of the diffusion coefficient $D^*$ obtained when neglecting the dielectric layer contribution ($\lambda_p = 0$), resulted to an overestimation of $D$ by more than three times.

\subsection{Transient currents}
\label{sec:exp:currents}
To verify the consistency of the model and the fitted parameters, the ion parameters were determined from measurements of transient currents in a cell under the voltage being suddenly applied to the electrodes.
In the Debye-Huckel approximation  \cite{Bazant:2004:PRE,Beunis:2007:APL}, small voltage $U_0$ applied to the electrolyte layer bounded by blocking electrodes with dielectric layers will lead to the initial current jump $I_0 =  U_0 / R$ with a subsequent exponential relaxation
\begin{equation}
\label{eq:cur}
 I = I_0 e^{ - t / \tau_s}  ,
\end{equation} 
where the characteristic time  $\tau_s$ is the same as appeared above in the frequency analysis of the impedance behavior, and is defined by Eq.~(\ref{eq:taus}).

The transient currents were measured in the cell with thickness $d = 10.3$~$\mu$m for several applied voltages $U_0$ and then fitted using Eq.~(\ref{eq:cur}).
\begin{figure}
\centering
\includegraphics[width=\figsize]{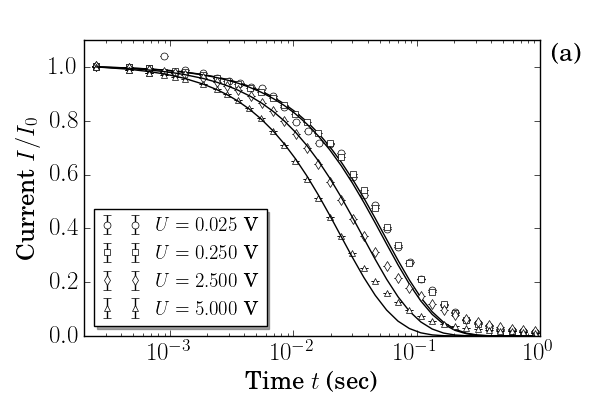}
\includegraphics[width=\figsize]{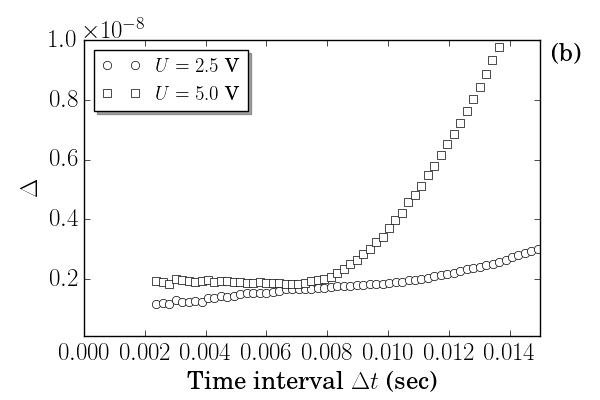}
\caption{Time dependence of the transient currents for the cell with thickness $d = 10.3$ $\mu$m at several applied voltages (symbols -- experimental data, solid lines -- fitted curves) (a). 
The mean-square deviation of experimental data on a linear dependence calculated for different fitting intervals (b).}
 \label{fig:pi-32-cur}
\end{figure}
Figure~\ref{fig:pi-32-cur} shows, that for the voltages $U_0 = 0.025$~V and $U_0 = 0.25$~V  the relative transient currents $I(t) / I_0$ are practically the same and well described by Eq.~(\ref{eq:cur}) in the time interval $t < 0.2$~s.
For higher voltages, $U_0 = 2.5$~V and $U_0 = 5.0$~V, the currents also relax exponentially, but with smaller characteristic times.
Note, that the deviation between the fitted curves and the experimental data at larger times $t > 0.2$~s can be attributed to the mechanisms responsible for the low-frequency behavior of the impedance, that was not taken into account in derivation of Eq.~(\ref{eq:cur}) \cite{Bazant:2004:PRE,Beunis:2007:APL}.

For each applied voltage $U_0$ the parameters $R$ and $\tau_s$ were determined by the least-square method and the charge relaxation time was calculated from $\tau_q = R\,C$, were the cell capacitance $C$ was separately measured by the RLC-meter.
\begin{table}
\caption{ \label{tab:pi-32-cur}Fitted parameters of transient currents for the cell with the thickness $d = 10.3$ $\mu$m.}
\begin{center}
\begin{ruledtabular}
 \begin{tabular}{*4c}
$U_0$, V &   $R$, M$\Omega$   &  $\tau_s$, ms & $\tau_q$, ms \\\hline
0.025 & 2.84 & 55.4 & 4.82 \\\hline
0.250 & 2.92 & 52.8 & 4.97 \\\hline
2.500 & 2.94 & 36.4 & 4.99 \\\hline
5.000 & 3.20 & 24.2 & 5.44 \\
\end{tabular}
\end{ruledtabular}
 \end{center}
 \end{table}
Comparison of Tables~\ref{tab:fitpars-fresh} and \ref{tab:pi-32-cur} shows, that for the voltages $U_0= 0.025$~V and $U_0= 0.25$~V the characteristic times $\tau_s$ and $\tau_q$ are very close to the values obtained from the analysis of the impedance spectra.
Hence, using Eqs.~(\ref{eq:lamdp}), (\ref{eq:DR}) will result in the ion parameters similar to those listed in Table~\ref{tab:ionpars-fresh} for the cell with $d = 10.3$~$\mu$m.

The values of $\tau_s$  obtained for the high applied voltages $U_0=2.5$~V and $U_0 = 5.0$~V are noticeably smaller than for the lower voltages that indicates a violation of the Debye-Huckel approximation.
In the limiting case of high applied voltage and small concentration of the ions an alternative approach considered in Ref.~\cite{Beunis:2008:PRE} can be used.
This approximation assumes the ion migration as only relevant mechanism of the charge transport under the electric field, as a consequence, the  electrical current should depend linearly on the time after the initial current jump
\begin{equation}
\label{eq:cur:max}
 I = I_0 (1 - t / \tau_{tr}) .
\end{equation} 
Here the initial current $I_0  $ and the characteristic time of the electrodiffusion $\tau_{tr}$ are given as
\begin{eqnarray}
\label{eq:I0:tautr}
I_0 = \frac{U_0}{R} =  \frac{2\, c_0 \hat{z}\, e\,  D\, S}{d} \frac{U_0}{U_T}, \quad
\tau_{tr} =\frac{d^2}{D}  \frac{U_0}{U_T} .
\end{eqnarray}
Extracting $I_0$ and $\tau_{tr}$ from the fit of the experimental data, the diffusion coefficient and the concentration of the ions can be calculated from 
\begin{equation}
\label{eq:cur:pars}
 D =  \frac{d^2}{\tau_{tr}} \frac{U_T}{U_0}, \quad c_0 = \frac{I_0}{2 \hat{z} e D} \frac{d}{S}  \frac{U_T}{U_0} = \frac{I_0}{2 \hat{z} e S d} \tau_{tr}.
\end{equation} 

In the course of time the ion movement under applied dc electric field will lead to gradual accumulation of the electric charges at the blocking electrodes and screening of the local electric field, that will result in the raising of the diffusion current.
Therefore, the linear regime given by Eq.~(\ref{eq:cur:max}) can only be observed for a certain initial time interval $\Delta t$ after the switching on the voltage, and its duration will depend on the ion concentration and voltage magnitude.

To determine the time interval $\Delta t$, where the linear regime holds, the data for  $U_0 = 2.5$~V and $U_0 = 5$~V were fitted using  Eq.~(\ref{eq:cur:max}) and mean-square deviations of the experimental data from the fitted curves  $\Delta$ were calculated  over the different initial time intervals $\Delta t$ in the range between 2~ms and 15~ms  (Fig.~\ref{fig:pi-32-cur}b).
For $U_0 = 2.5$~V the values of mean-square deviations $\Delta$ are continuously changed over the whole range of $\Delta t$, in contrast to the case  $U_0 = 5$~V, where $\Delta$ remains almost constant within time $\Delta t < 7$~ms that implies a linearity of the current over this time interval.
Using the fitted parameters $\tau_{tr} = 28$~ms and $I_0 = 1.55$~$\mu$A corresponding to $\Delta t = 7$~ms, the diffusion coefficient  $D = 19$ $\mu$m$^2$/s and  the ions concentration $c_0 = 119$~$\mu$m$^{-3}$ have been obtained from Eq.~(\ref{eq:cur:pars}). 
These values of $D$ and $c_0$ are close to the values obtained from the analysis of the impedance spectra (see Table~\ref{tab:ionpars-fresh}).

Thus, the analysis of the transient currents in cases of low and high applied voltages confirms the validity of the results obtained from the analysis of the impedance spectra based on the model of the blocking electrodes with the dielectric layers.

\subsection{Impedance behavior with LC cell aging}
\begin{figure}
\centering
\includegraphics[width=\figsize]{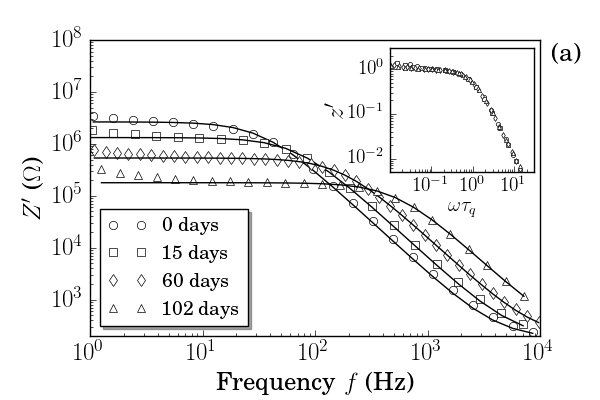}
\includegraphics[width=\figsize]{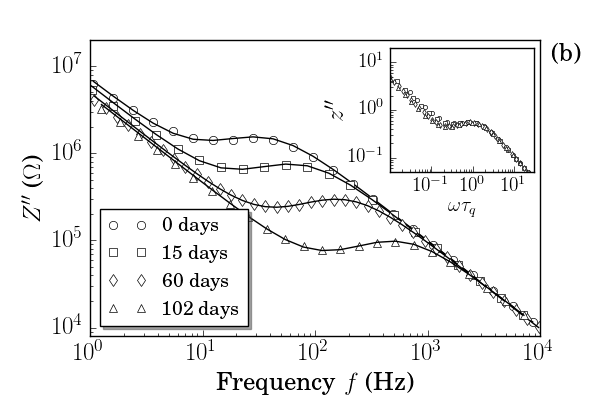}
\includegraphics[width=\figsize]{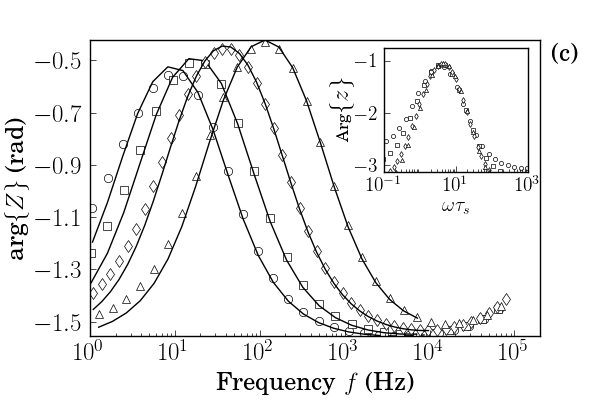} 
\caption{Impedance spectra of the LC cell with the thickness of $d = 10.3$~$\mu$m measured over long time intervals. Symbols are experimental data and solid lines are fitting curves. Insets in parts (a) and (b) show the dimensionless impedance dependence on $\omega \tau_q$ and in part (c) the argument $\arg{Z}$ normalized by its maximum value as a function of $\omega \tau_s$.}
 \label{fig:imp-aging}
\end{figure}

Earlier studies showed that the conductivity of LC cells filled with the liquid crystal 5CB exponentially increased with time with two characteristic time scales $\sim 400$~h and $\sim 4000$~h \cite{Naito:1995:MCLC, Mada:1996:JJAP, Murakami:1997:JJAP}.
To investigate this phenomenon the impedance spectra in the cell with the thickness $d=10.3$ $\mu$m were measured after a while over three months (Fig.~\ref{fig:imp-aging}).
%
The cell was not completely sealed and was stored in a dark place at the  temperature $T=25^{\circ}$C and the relative humidity $(70 \pm 10)$\% between measurements.
Figure~\ref{fig:imp-aging} demonstrates that the plateau level in the real part of the impedance $Z'$ decreases with aging time reflecting an increase in the conductivity of the LC cell.
At high frequencies the values of the imaginary part $Z''$ remain practically unchanged, but the extrema values are decreased and their positions are shifted towards higher frequencies.
The maximum of the argument $\arg{Z}$ increases and shifts to the high frequency range.
The insets in Figs.~\ref{fig:imp-aging}(a), (b) show the dependence of the dimensionless impedance $z =Z/R$ on dimensionless frequency $\omega \tau_q$ and demonstrate a frequency scaling with respect to the conductivity increase with aging time.
The frequency scaling is also held near the maximum of the impedance argument [Fig.~\ref{fig:imp-aging}(c)], if $\tau_s$ is chosen as a relevant time scale.

\begin{figure}
\centering
\includegraphics[width=\figsize]{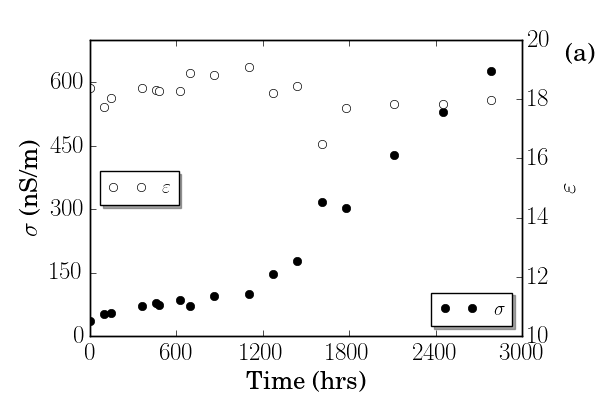}
\includegraphics[width=\figsize]{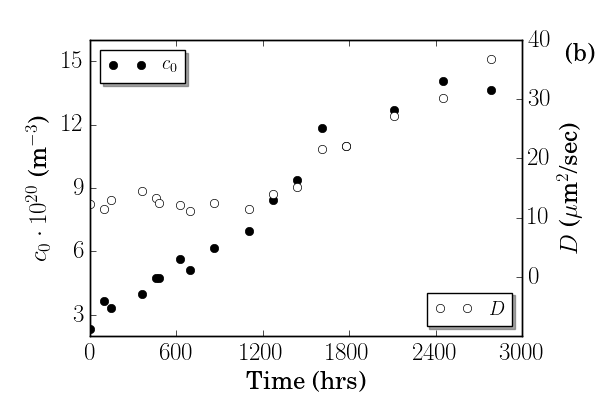}
\includegraphics[width=\figsize]{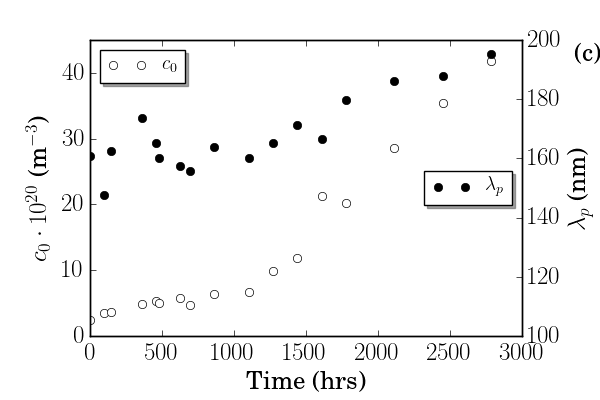} 
\caption{Time dependence of the conductivity $\sigma$ and the dielectric constant $\varepsilon$ (a), the concentration of the ions $c_0$ and the diffusion coefficient $D$ (b),  the concentration of the ions $c_0$ and the effective  thickness of the dielectric layer   $\lambda_p$ (c). 
The values in (b) were calculated for the fixed value of  $\lambda_p =150$~nm, and in (c) for the fixed value of the diffusion coefficient $D = 12$~$\mu$m$^2$/s.}
 \label{fig:imp-aging-pars}
\end{figure}

An existence of the scaling in the experimental data with respect to the conductivity change indicates that the frequency dependence of the impedance can be described by the proposed model.
The unchanged imaginary part of the impedance $Z''$ at high frequencies, simultaneously with the decrease in the plateau level of the real part $Z'$ and shifting of $Z''$ maximum to the higher frequencies, point out that the charge relaxation time $\tau_q$ decreases with aging time only due to the conductivity increase, which depends on the concentration and the diffusion coefficient of the ions.
On the other hand, the shifting of the minimum of $Z''$ and the maximum of  $\arg{Z}$ to the high frequency range can also be caused by a change in the effective thickness of the dielectric layers.
The time dependence of the conductivity $\sigma$, the dielectric constant $\varepsilon$, the concentration of the ions $c_0$ and the diffusion coefficient $D$ calculated from the fitted parameters of the impedance spectra for the fixed value of the effective thickness of the polymer layer $\lambda_p =150 $~nm are shown in Figs.~\ref{fig:imp-aging-pars}(a), (b).

One finds the dielectric constant is almost unchanged over the aging time [Fig.~\ref{fig:imp-aging-pars}(a)].
The conductivity changes with time over the two different ranges: in the interval $t < 1200$~h  the values of $\sigma$ increase about twice (from 50~nS/m to 100~nS/m), whereas for $t > 1200$~h $\sigma$ increases by six times (from 100~nS/m up to 600~nS/m).
The concentration of the ions $c_0$ practically linear increases with aging time [Fig.~\ref{fig:imp-aging-pars}(b)].
The diffusion coefficient remains approximately constant for $t <1200$~h and is equal to $D = (12.4 \pm 2.8)$~$\mu$m$^2$/sec as calculated by averaging over this time interval.
For $t >1200$~h the value of $D$ increases by about three times, that  can be interpreted as an appearance in the liquid crystal significant fraction of the ``fast '' ions with small Stokes radius.

On the other hand, assuming that the ions remain of the same type, therefore, the diffusion coefficient is unchanged, one can find the Debye length from $\lambda_D =  \sqrt{\tau_q D}$ and calculate the effective thickness of the polymer layer $\lambda_p$ from Eq.~(\ref{eq:lamdp}) using the values of $\tau_q$ and $\tau_s$ obtained from the fitting of the impedance spectra.
Figure~\ref{fig:imp-aging-pars}(c) shows corresponding time dependence of the concentration of the ions $c_0$ and the effective thickness of the dielectric layer $\lambda_p$ calculated for the fixed value of the diffusion coefficient $D = 12$~$\mu$m$^2$/s.
In this case, the behavior of $c_0(t)$ becomes similar to the dependence of $\sigma(t)$ in Fig.~\ref{fig:imp-aging-pars}(a), and the value of $c_0$ at $t =3000$~h is three times larger than that obtained for the case of the fixed effective thickness of the dielectric layer  $\lambda_p =150$~nm.
According to Fig.~\ref{fig:imp-aging-pars}(c), $\lambda_p$ remains constant for $t <1200$~h and for $t >1200$~h increases by about $45$~nm.
The temporal change of $\lambda_p$ obtained under an assumption of the unchanged diffusion coefficient can be interpreted either as a formation of adsorbed ionic layers at the boundaries between the polymer film and the LC layer, or a swelling of the polymer film caused by the diffusion of the LC molecules into the polymer.
In the latter case an increase in $\lambda_p$ from $150$~nm to $195$~nm corresponds to the growth of the thickness of the polymer film from $30$~nm to $39$~nm ($h=\varepsilon_p \lambda_p / \varepsilon$).

Thus, the measurements of the impedance spectra shows that during the aging of the liquid crystal cell the dielectric constant is not changed over $3000$~h after the cell filling.
For the same time interval, the conductivity was increased by one order of magnitude.
Based on the proposed model we can conclude that the conductivity growth over $1200$~h after the cell filling is originated from the increase of the concentration of the same type of ions as in the fresh made cell.
It is proved by the constant value of the diffusion coefficient and effective thickness of the polymer layer found over this time interval. 
However, for longer time of cell aging the model leads to ambiguous estimates of the ion parameters.
Therefore, to identify physical processes at longer time, more complicated theoretical models considering such effects as  polymer swelling or long-term ion adsorption have to be developed.

\section{Discussion}
\label{sec:disc}
\begin{figure}
\centering
\includegraphics[width=\figsize]{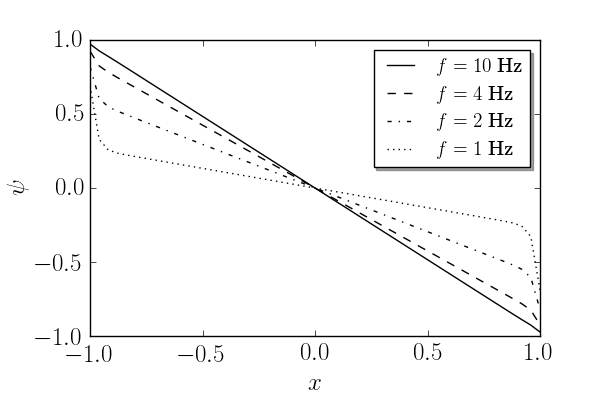}
\caption{The distribution of the electric potential for different frequencies for the cell with the thickness $d = 17.4$~$\mu$m.}
 \label{fig:imp-psi}
\end{figure}

The values of the diffusion coefficient of the ions in the liquid crystal 5CB in the range $D = (6 - 10)$~$\mu$m$^2$/sec were reported in studies of transient currents under polarity reversal of the applied dc voltage~\cite{Sugimura:1991:PRB, *Sugimura:1990:MCLC} and currents induced by UV-light in LC cells with photosensitive semiconducting electrodes~\cite{Murakami:1997:JJAP}. 
The diffusion coefficients of the ions in 5CB were also estimated from the analysis of the dielectric spectra in the framework of the constant electric field model already mentioned in the Introduction.
Considering dielectric behavior of binary electrolyte under the applied dc bias it has been shown that ions concentration and diffusion coefficient can be determined from the high frequency part of the dielectric spectra~\cite{Uemura:1972:JOPSPPE, *Uemura:1974:JOPSPPE}.
Based on this model, the temperature dependence of the diffusion coefficient of the ions in 5CB was determined from the dielectric spectra of the LC cells with and without polyimide layers on top of the ITO electrodes~\cite{Naito:1995:MCLC, Murakami:1997:JJAP}.
For both samples the diffusion coefficients were found nearly the same $D \approx 16$~$\mu$m$^2$/sec at $T=23^\circ$C.
In another version of the constant electric field model~\cite{Sawada:1998:MCLC, *Sawada:1999:MCLC} a presence of different kinds of ions was considered and an additional parallel RC-circuit to the expression for the total complex dielectric constant was included to take into account possible surface effects, such as electric double layers on the electrodes.
The best fit of the experimental dielectric spectra of 5CB in the wide frequency range $(10^{-3}-10^{3})$ was obtained using $5$ kinds of ions with the diffusion coefficient of the average kind of ions $D = 4.9$~$\mu$m$^2$/sec at $T=23^\circ$C \cite{Naemura:2003:MCLC}.
Note, that all results mentioned above were obtained for LC cells with homogeneous planar orientation of the liquid crystal and the estimates of the diffusion coefficient are related to the perpendicular component $D_{\bot}$.
In present work the homeotropic layers of 5CB were studied and the parallel component $D_{||} = (12.4 \pm 2.8)$~$\mu$m$^2$/sec was determined.
Taking into account the anisotropy of the conductivity of 5CB $\sigma_{||} / \sigma_\bot = 1.65$ \cite{Naemura:2003:MCLC} the perpendicular component $D_\bot = (7.5 \pm 1.7) $~$\mu$m$^2$/sec can be calculated, that is in a good agreement with the diffusion coefficients estimated from the transient and photo-induced currents and dielectric spectra analysis based on the constant electric field model.
The constant electric field model is able to describe the impedance (dielectric) spectra over a wider frequency range than one based on the PNP equations.
However, justifications for the constant electric field model are the subject of lengthy debates \cite{Alexe-Ionescu:2009:PRE, Sawada:2013:PRE, Alexe-Ionescu:2014:JCP, Sawada:2016:PRE}.
To understand possible reasons for closed values obtained from the constant electric field approach and in this work, we have calculated from Eqs.~(\ref{eq:psi:sol}), (\ref{eq:psi0}) the distribution of the electric potential  $\psi(x)$ for different frequencies using the parameters of the LC cell with the thickness $d = 17.4$~$\mu$m (Fig.~\ref{fig:imp-psi}).
Figure~\ref{fig:imp-fresh} shows that for this cell the maximum of the impedance argument is observed at $f \approx 10$~Hz.
As it follows from Fig.~\ref{fig:imp-psi}, at this frequency the distribution of the electric potential across the cell is almost linear (electric field is a constant).
Therefore, in this case the constant electric field approximation is held and should lead to the similar value of the diffusion coefficient as obtained here.
Strong deviations of $\psi(x)$ from the linear dependence for the frequencies $f < 10$~Hz on Fig.~\ref{fig:imp-psi}) indicate violation of the constant electric field assumption at low frequencies.
This makes uncertain the estimates of the parameters of different kinds of ions from the fitting of the low frequency part of the impedance (dielectric) spectra in the framework of the constant field approach (see also discussion in \cite{Alexe-Ionescu:2009:PRE}).

Unfortunately, in the framework of the PNP approach, there are no estimates of the diffusion coefficient of the ions in the liquid crystal 5CB from the analysis of the impedance or dielectric spectra.
More sophisticated models for the binary electrolyte with nonblocking electrodes where the currents are generated by the ion adsorption/desorption processes have been considered in Refs.~\cite{Alexe-Ionescu:2009:JAP, Paula:2012:PRE}.
The expressions for the impedance derived there were used to fit the experimental spectra of 5CB in the cells with polymer or silicone oxide coated electrodes choosing values of the diffusion coefficients  $D = 1.9$~$\mu$m$^2$/sec \cite{Alexe-Ionescu:2009:JAP} and $D = 2.5$~$\mu$m$^2$/sec \cite{Paula:2012:PRE}.
An agreement between theory and experiment was found to be good in the limited range of frequencies, but the influence of the chosen values of $D$ on the resulting values of the ions concentration and parameters of adsorbing currents was not discussed.

We have demonstrated (see Table~\ref{tab:ionpars-fresh}), that neglecting the dielectric layer in the estimate of the Debay length results in about threefold increase in the value of the diffusion coefficient.
Similar reason may explain the contradictory values of the ion parameters obtained by means of the pulsed-field gradient NMR method and on the basis of the impedance spectra analysis \cite{Wang:2013:PRE}.
According to Eq.~(\ref{eq:DR}), using $\lambda_s$ instead of $\lambda_D$ in systems with $\lambda_p \sim \lambda_D$ will lead to overestimated diffusion coefficient and underestimated charge carrier concentration, that corresponds to the results in Ref.~\cite{Wang:2013:PRE}.
This is especially true for highly concentrated electrolytes, where the Debye length $\lambda_D$ is known to be small and thus comparable to the effective thickness $\lambda_p$ of compact layers formed by the adsorbed ions.
The universal behavior of the impedance spectra is inherent for ionic conductors of different nature, such as ion glasses, polymer electrolytes, ionic liquids or aqueous solutions of salts \cite{Dyre:2000:RMP, Sangoro:2008:PRE, Serghei:2009:PRB}.
In the high frequency range the impedance spectra can be scaled relative to the ion concentration, temperature and sample thickness, but at low frequencies the scaling fails when the samples thickness or the electrode materials are changed.
The dimensionless expressions for the impedance Eq.~(\ref{eq:imp:exact}), (\ref{eq:imp:sim}) clearly show, that no single characteristic time (or frequency) can be selected to scale the spectra in the whole frequency range.
As it follows from the analysis of the approximate expression Eq.~(\ref{eq:imp:sim}), at high frequencies, $\omega \tau_q \gg 1$, the behavior of the impedance spectra is determined by the dynamics of the bulk charges with the characteristic time $\tau_q$.
The value of $\tau_q$ depends only on the charge carrier properties -- the concentration of the ions and the diffusion coefficient, resulting in the observed scaling of the impedance spectra  $z = f(\omega \tau_q)$ in the high-frequency range  when changing the ion concentration, temperature and sample thickness \cite{Dyre:2000:RMP, Sangoro:2008:PRE, Serghei:2009:PRB}.
Our results also confirm the high-frequency scaling with respect to the thickness of the liquid crystal layer and its conductivity [Figs.~\ref{fig:imp-aging}(a), (b)].
At low frequencies, $\omega \tau_q \ll 1$, the behavior of the impedance is dominated by the surface effects: the diffuse layer dynamics and the dielectric layer charging, that in the case of $\lambda_D \ll L$ act together as a single nonconducting dielectric layer with characteristic charging time $\tau_s$.
In the case studied, the scaling of the impedance also holds in the low-frequency range, if $\omega \tau_s$ is chosen as the dimensionless frequency  [Fig.~\ref{fig:imp-aging}(c)].
The characteristic time $\tau_s$ is determined by the distance between the electrodes, the properties of the charge carriers and the dielectric layers [Eq.~(\ref{eq:taus})].
It can be assumed, that the scaling $z = f(\omega \tau_s)$ can also hold in the low-frequency range when the electrode materials are changed \cite{Sangoro:2008:PRE, Serghei:2009:PRB} due to existence of the compact layers, whose properties are defined by the electrode adsorption ability determining the effective thickness and, hence, the value of $\tau_s$.

\section{Conclusion}
\label{sec:conc}

In this work, the impact of the dielectric layers on the frequency dependence of the impedance of ionic conductors have been investigated.
In the framework of the PNP approach the expression for the impedance of the symmetric binary electrolyte with blocking electrodes and dielectric layers has been derived and the frequency dependence has been expressed in terms of contributions with three characteristic times in the system -- relaxation of the bulk charge $\tau_q$ and charging times of the diffuse $\tau_c$ and dielectric $\tau_p$ layers.
The frequency dependence of each contribution is different, that suggests the values of all characteristic times can be determined from the analysis of the impedance spectra and then the ion parameters and the effective thickness of the dielectric layer can be calculated.
However, the analysis demonstrates, that in most common cases of thin diffuse layer, $\lambda_D \ll L$,  only the sum of the Debye length and the effective thickness of the dielectric layer,  $\lambda_s = \lambda_D + \lambda_p$, can be obtained from the impedance spectra.
The established relations between the ion parameters and the characteristic times show that in the case of comparable thickness $\lambda_p \sim \lambda_D$, the correct estimates of the ion parameters from the impedance spectra is only possible if the properties of the dielectric layers are known.
Note, that the properties of the ions and of the dielectric layers can be simultaneously determined from the impedance spectra using the exact expression for the impedance, but only in systems with $\lambda_D \sim L$, such as nanochannels and porous media.
To verify considered model the system with comparable values of the Debay length and the effective thickness of the dielectric layers was experimentally studied.
The impedance spectra of the liquid crystal 5CB in the cells with different thickness containing electrodes coated by thin polymer films with known properties were measured.
It was shown, that exact and approximate expressions for the impedance fit data almost equally well and give close values for the ion parameters for thick cells, whereas for thin cells the exact expression fits experimental data better.
For all cells used with the thickness in the range $(3 - 17)$~$\mu$m the Debay length $\lambda_D \approx 200$~nm and the diffusion coefficient $D \approx 12$~$\mu$m$^2$/s have been determined. which remains unchanged over 1200~h of the cell aging.
Similar value of the diffusion coefficient was obtained from the measurements of the transient currents in low and high limits of the applied dc voltage.
Corresponding hydrodynamic radius of the ions $R_g \approx 0.3$~nm is close to the radius of free inorganic ions \cite{Israelachvili:2011}.
This supports an assumption that the conductivity of liquid crystals is caused by free inorganic impurity ions left after a synthesis or emerging from materials used to assemble LC cells.

However, correct estimations of the ion parameters were only possible for known dielectric constant and thickness of the polymer layers at the electrodes and for relatively fresh samples.
For long time of cell aging above 1200~h, the estimates of the ion parameters become ambiguous that allows different interpretations.
In this case, it can not be unequivocally concluded how an ionic composition of the liquid crystal and/or the properties of the polymer film are changing with time.
%
%
In the systems, where an electrolyte is in contact with metallic electrodes, the situation can be much worse, since the properties of the compact layers formed at the electrode surfaces are quite difficult to characterize.
Thus, the analysis of the impedance spectra in such systems based on the considered model should be carried out with caution, especially in the case of highly concentrated electrolytes.

\begin{acknowledgments}
We thank E.~S.~Batyrshin and Yu.~S.~Zamula for measurements of the polymer layer thickness by the AFM technique and A.~P.~Krekhov for stimulating discussions and critical reading of the manuscript.
\end{acknowledgments}


%

\end{document}